\documentclass[12pt]{article}

\usepackage{times}
\usepackage{graphicx}
\usepackage{proofread}
\usepackage{lineno} 
\usepackage{caption}
\usepackage{capt-of}

\title{Controlling the Photoelectric Effect in the Time Domain} 

\author{Yu-Chen Cheng,$^{1\dagger}$ Sara Mikaelsson,$^{1\dagger}$ Saikat Nandi,$^{1}$ Lisa R\"amisch,$^{1}$ Chen Guo,$^{1}$ \\
Stefanos Carlstr\"om,$^{1}$ Anne Harth,$^{1}$ Jan Vogelsang,$^{1}$ Miguel Miranda,$^{1}$ \\
Cord L. Arnold,$^{1}$ Anne L'Huillier,$^{1}$ and Mathieu Gisselbrecht$^{1\ast}$ \\
\\
\normalsize{$^{1}$Department of Physics, Lund University, P.O. Box 118, 22100 Lund, Sweden}\\
\\
\normalsize{$^\dagger$These authors contributed equally to this work}
\\
\normalsize{$^\ast$E-mail:  mathieu.gisselbrecht@sljus.lu.se}}

\date{}

\begin{document}

\baselineskip 24pt

\maketitle 

\renewcommand{\abstractname}{\vspace{-\baselineskip}} 

\begin{abstract}
\textbf{
When small quantum systems, atoms or molecules, absorb a high-energy photon, electrons are emitted with a well-defined energy and a highly-symmetric angular distribution, ruled by energy quantization and parity conservation.
These rules seemingly break down when these systems are exposed to short and intense optical pulses, which raise the question of their universality for the simplest case of the photoelectric effect. 
Here we investigate photoionization of helium by a sequence of attosecond pulses in the presence of a weak infrared laser field. We continuously control the energy and introduce an asymmetry in the emission direction of the photoelectrons, in apparent contradiction with the quantum-mechanical predictions. This control, made possible by the extreme temporal confinement of the light-matter interaction, opens a new road in attosecond science, namely, the manipulation of ultrafast processes with a tailored sequence of attosecond pulses.}
\end{abstract}

\paragraph*{Introduction}

Since the seminal scientific contributions of Planck \cite{PlanckAdP1901} and Einstein \cite{EinsteinAdP1905} at the beginning of the 20\textsuperscript{th} century, it is well known that matter absorbs light in the form of discrete energy quanta ($h\nu$, the photon), where $h$ is the Planck constant and $\nu$ is the light frequency. Photoabsorption in centrosymmetric systems such as free atoms or molecules follows strict selection rules\footnote{described within the dipole approximation}, with a change of parity between the initial and final states \cite{LaporteJOSA1925}. When the absorbed energy is above the binding energy ($I_p$), a photoelectron is emitted with kinetic energy equal to $h\nu-I_p$ \cite{EinsteinAdP1905} and its probability of emission is symmetric relative to the origin \cite{YangPR1948,CooperJCP1968}.
With the advent of bright monochromatic light sources such as lasers \cite{MaimanNature1960} and synchrotron radiation sources \cite{RowePA1973} as well as the progress in photoelectron detection technology, in-depth studies of photoemission in a variety of systems with ever increasing energy and angle resolution have confirmed these quantum mechanical predictions \cite{QuackHandbook2011,BeckerVUV2012}.

As the light intensity increases, non-linear multiphoton processes become possible, leading to new ionization mechanisms. In above-threshold-ionization (ATI) processes, electrons are emitted  at discrete kinetic energies \cite{AgostiniPRL1979,BucksbaumPRL1987} and angular distributions remain symmetric, except in some particular multiphoton schemes using several frequencies that mix parity in the final state \cite{YinPRL1992,LaurentPRL2012}.
This picture breaks down in strong and ultrashort infrared (IR) laser fields, when ionization essentially takes place during less than an optical cycle. As shown in the ``stereo ATI'' technique \cite{PaulusNature2001,PaulusPRL2003}, atoms exposed to intense few-cycle pulses emit electrons  with a continuous kinetic energy distribution and a small asymmetry with respect to the origin which depends on the carrier-to-envelope (CEP) phase offset of the laser pulses.   

With even shorter pulses, produced through high-harmonic generation in gases \cite{McPhersonJOSAB1987,FerrayJPB1988} in the attosecond extreme ultraviolet (XUV) range, new tools become available for the study time-resolved photoemission processes in atoms \cite{SchultzeScience2010,  KlunderPRL2011,OssianderNatPhys2017,IsingerScience2017}, molecules \cite{HuppertPRL2016, CattaneoNatPhys2018,VosScience2018} and solids \cite{CavalieriNature2007,LucchiniScience2016}.
The ``streaking'' technique \cite{KienbergerScience2002,PazourekRMP2015} combines a single attosecond pulse with an intense IR laser pulse. In this case, the kinetic energy distribution of the photoelectrons, imposed by the attosecond pulse bandwidth, is very broad, typically several eV, and can be continuously varied depending on the delay between the XUV and IR fields. The energy shift can be understood classically by momentum transfer from the IR electromagnetic field to the electron which is released by absorption of an XUV photon. 
At the delays when the energy transfer is not zero, the angular distribution is asymmetric.
The reconstruction of attosecond harmonic beating by interference of two-photon transitions (RABBIT) technique \cite{VeniardPRA1996, PaulScience2001,MullerAPB2002} uses a train of attosecond pulses together with a weak IR laser pulse. In this case, the photoelectron momentum distribution remains symmetric and the kinetic energy spectrum presents discrete peaks separated by the IR photon energy.

This brief description of the state of knowledge of light-matter interaction shows that the rules of energy quantification and parity conservation, established at the beginning of last century for describing the photoelectric effect, are not universal, in particular for short and intense optical fields. To the best of our knowledge, the limits of these rules in photoelectron spectroscopy, in particular concerning energy quantification, have not been discussed in the case of weak optical fields.

\begin{figure}[ht]
\includegraphics[width=1\textwidth]{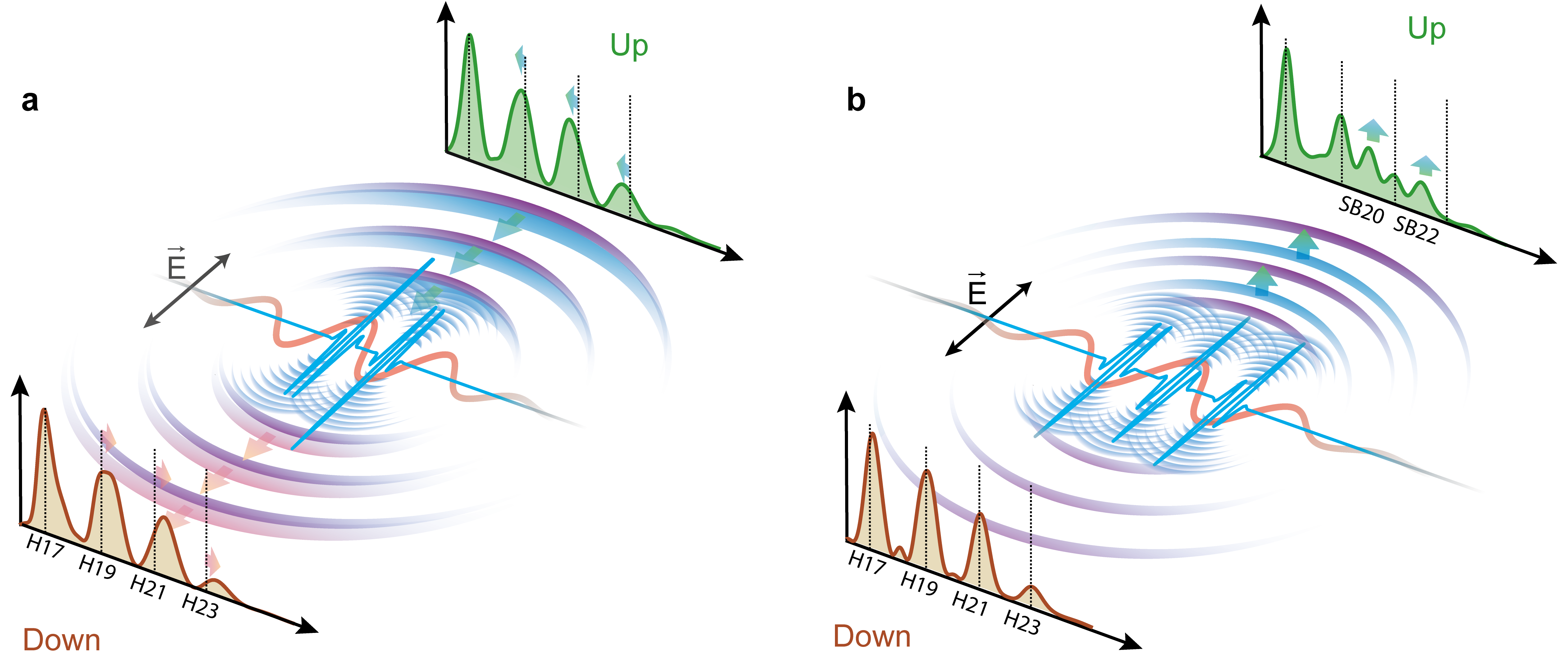}\\
\caption{\textbf{Principle of the experiment:} Helium atoms are exposed to two (\textbf{a}) or three (\textbf{b}) XUV attosecond pulses (blue) in presence of a weak IR laser field (red) at a fixed delay. Electron wave packets (violet) are emitted  with an up-down asymmetry relative to the direction of polarization, resulting in different spectra (brown and green) when recording electrons emitted in the two opposite directions.
In the case of two pulses (\textbf{a}), the photoelectron spectrum is shifted towards higher or lower energies, while for three pulses (\textbf{b}), peaks at different frequencies, called sidebands, are observed, mostly in the up direction.}
\label{fig:principle}
\end{figure}

In this work, we examine the transition between a ``classical'' energy transfer from the IR electromagnetic field to the photoelectron, where  the photoelectron momentum can be changed continuously, and a ``quantum-mechanical'' picture of light-matter interaction, where the kinetic energy of the photoelectron varies by discrete quanta, equal to the absorbed/emitted IR photon energy.  
To this end, we study the photoionization of helium by tailored sequences of few attosecond pulses in presence of a weak IR laser field using three-dimensional momentum electron detection, which is becoming an essential tool in attosecond science \cite{HoglePRL2015,HeuserPRA2016}. 
 The principle of our experiment is illustrated in Fig.~\ref{fig:principle}. Helium atoms interact with two ({\bf a}) or three ({\bf b}) attosecond pulses,  and the IR field. Electron wavepackets are emitted, which carry the phase of the ionizing attosecond pulse and a phase modulation due to the IR field at the time of ionization. The resulting momentum distribution is determined by the interference of these wavepackets. 
When helium atoms are photoionized by two attosecond pulses separated by half the laser period, the electron energy is shifted relative to the kinetic energy for the XUV-only case by a {\it continuous} amount which depends on the IR light field as well as on the direction of emission. When helium atoms are photoionized by three attosecond pulses, we recover discrete energies equal to the energy of the absorbed photons minus the ionization energy. The emission direction is, however, strongly asymmetric. A theoretical analysis shows that this behavior can be explained by the extreme temporal confinement of the light-matter interaction to less than a cycle of the IR field, shedding light on the possibility to control the photoelectric effect in the time-domain.

\begin{figure}[ht]
\includegraphics[width=0.8\textwidth]{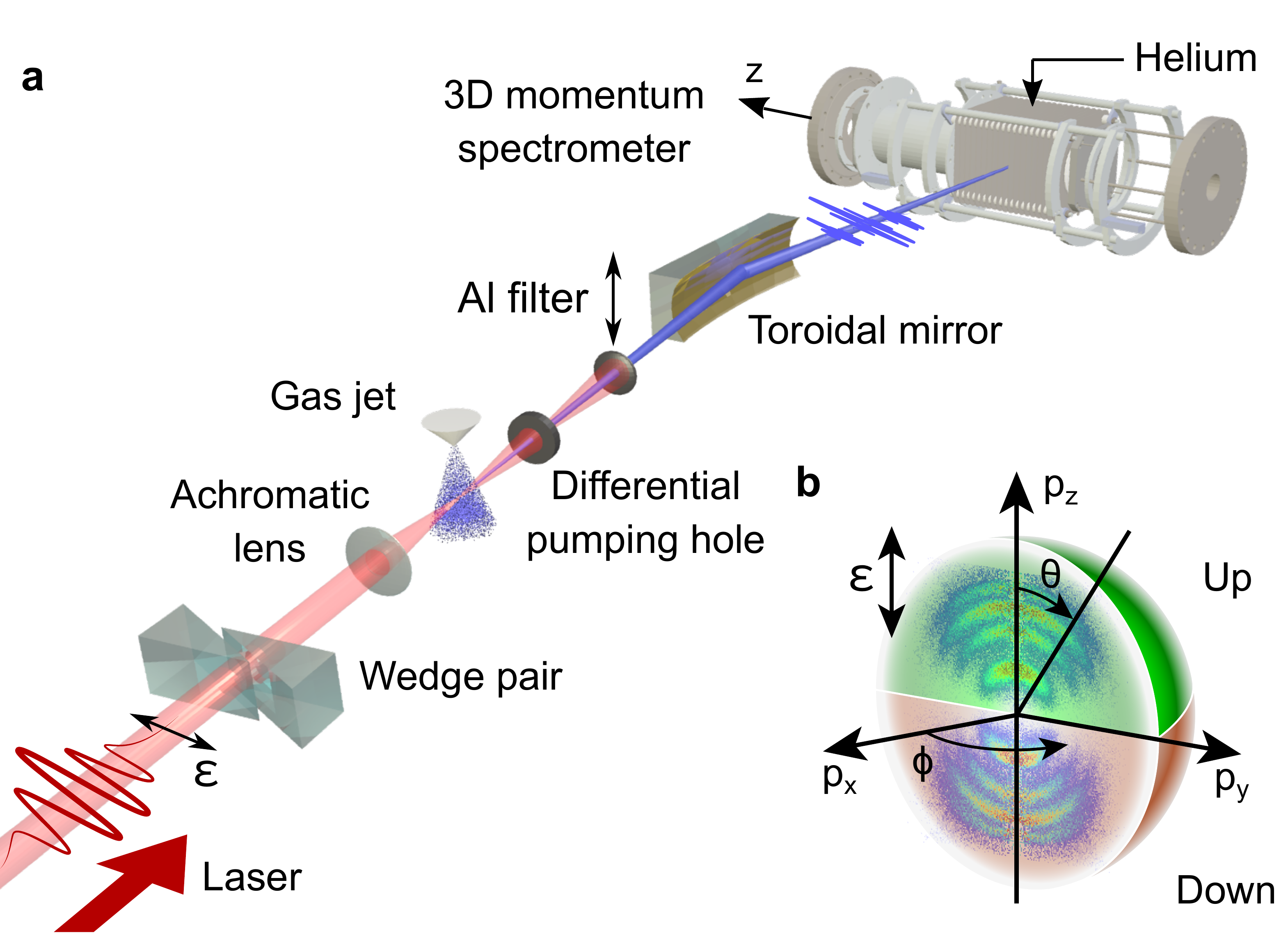}\\
\caption {\textbf{Experimental setup:} (\textbf{a}) 200-kHz 6-fs IR laser pulses with horizontal polarization are sent through a wedge pair for CEP control and focused with an achromatic lens into a high-pressure argon gas jet. A tailored sequence of few XUV attosecond pulses is then generated and focused by a gold-coated toroidal mirror into a 3D momentum spectrometer, where it intersects an effusive helium jet. An Al filter can be introduced to eliminate the co-propagating IR field. (\textbf{b}) 3D representation of electron momentum distribution as a function of azimuthal angle $\phi$ and angle $\theta$ for XUV-only radiation.}
\label{fig:experiment}
\end{figure}

\paragraph*{Results}
Details of the experiment are presented in Fig.~\ref{fig:experiment}~({\bf a}) and explained in the methods section. Briefly, a 200-kHz repetition-rate CEP-stable few-cycle laser generates few attosecond pulses separated by half the laser period (1.3~fs) \cite{GuoJPB2018}. Helium atoms are ionized by the attosecond pulses, in presence of a weak fraction of the IR laser pulse. In contrast to RABBIT or streaking experiments, the delay between the XUV and the IR fields is kept fixed. Charged particles are detected using a three-dimensional (3D) momentum  spectrometer based on an electron-ion coincidence scheme \cite{DornerPhysRep2000,Ullrich03rpp}. 
Fig.~\ref{fig:experiment}~({\bf b}) shows an example of a 3D photoelectron momentum distribution obtained in helium with XUV-only radiation. Since the momentum distribution has rotational symmetry along the polarization axis, we can define the photoelectron direction with positive (negative) $p_z$ as up (down). Simulations based on the Strong Field Approximation (SFA) \cite{LewensteinPRA1994,QuereJMO2005} are described in the methods section.

\begin{figure}[ht]
\includegraphics[width=0.8\textwidth]{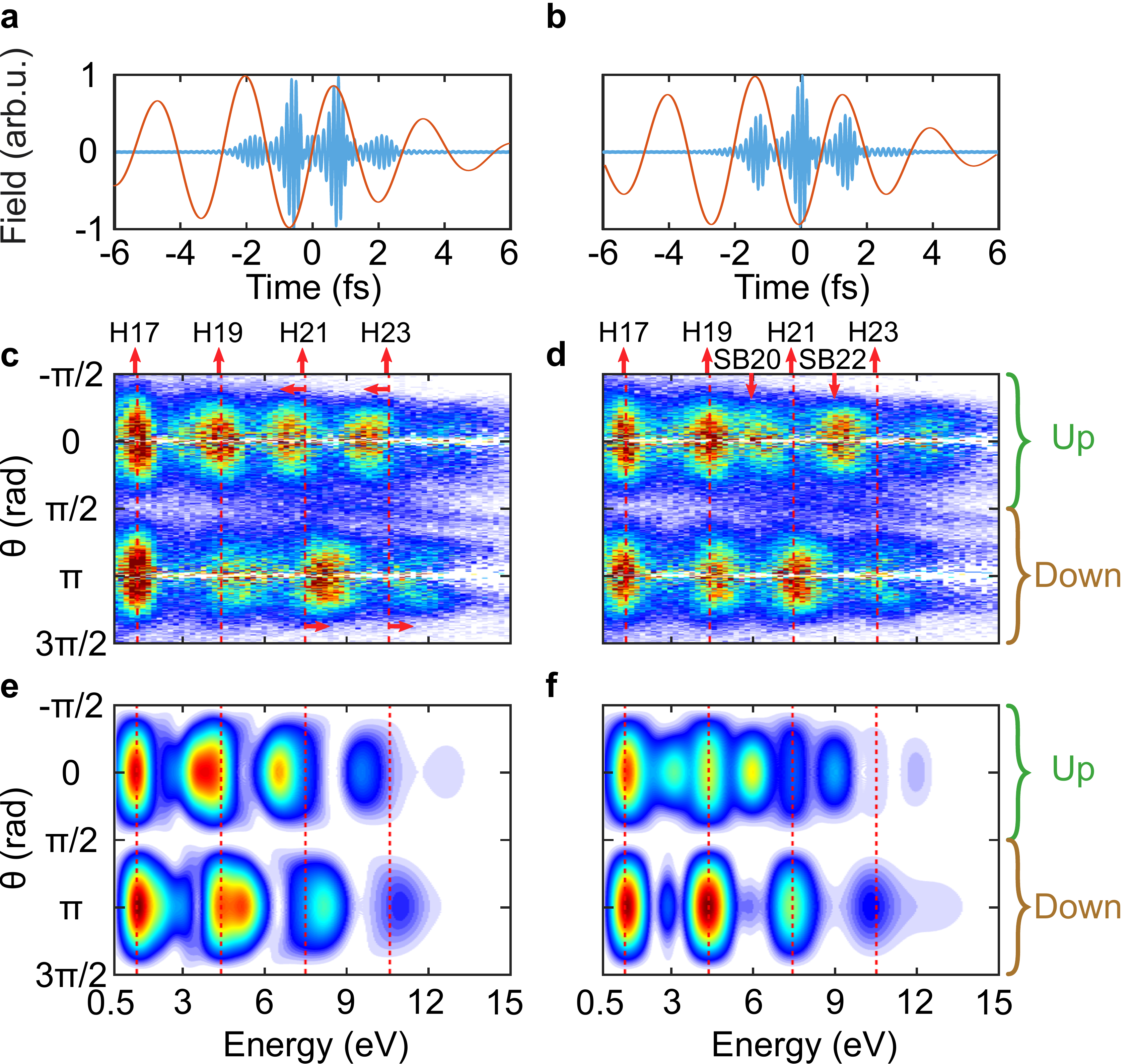}\\
\caption {\textbf{XUV attosecond pulse trains and angular-resolved spectrograms:} ({\bf a},{\bf b}) XUV (blue) and IR (red) electric fields. (\textbf{a}) $\pi/2$ and (\textbf{b}) 0; ({\bf c}-{\bf f}) Color plots representing the photoelectron angular distributions as function of energy. The experimental results are shown in ({\bf c},{\bf d}), while corresponding simulated photoelectron spectra are shown in ({\bf e},{\bf f}). The red dashed lines indicate the photoelectron kinetic energies after absorption of XUV radiation. When two attosecond pulses are used, the electron distribution shifts in energy, in opposite ways for the up and down emission directions ({\bf c},{\bf e}). In the three-pulse case, sidebands appear, but only in the up direction ({\bf d},{\bf f}). \\}
\label{fig:result}
\end{figure}

Fig.~\ref{fig:result} shows photoelectron distributions as a function of angle and energy, in two cases corresponding to attosecond pulse trains generated by IR fields with CEPs equal to $\pi/2$ ({\bf a}) and~0 ({\bf b}). Measured and simulated results are shown in ({\bf c}, {\bf d}) and ({\bf e}, {\bf f}), respectively. As shown in Fig.~3~({\bf a}), laser pulses with CEP equal to $\pi/2$, antisymmetric with respect to time reversal, lead to the generation of an even number of attosecond pulses. In our conditions, we obtain mainly two similar pulses of opposite sign, since they are generated by two consecutive half cycles of the IR field. The resulting photoelectron distributions, shown in Fig.~3~({\bf c}) and ({\bf e}), are shifted towards lower energy in the up direction and higher energy in the down direction. The shift increases with kinetic energy. 

As shown in Fig.~\ref{fig:result}~({\bf b}), laser pulses with CEP equal to $0$, symmetric with respect to time reversal, lead to the generation of an odd number of attosecond pulses, with a main central pulse. In our conditions, we obtain three pulses. The resulting photoelectron distributions, shown in Fig.~\ref{fig:result}~({\bf d}) and ({\bf f}), depend on the direction of emission. In the down direction, photoelectrons are emitted with kinetic energies corresponding to absorption of harmonics 17 to 23. In the up direction, photoabsorption of harmonics 21 and 23 is strongly reduced while new peaks corresponding to additional absorption/emission of an IR photon appear, so-called sidebands (SB20 and SB22). In general, measurements performed with different laser CEP show both energy shifts and asymmetric appearance of sidebands. When the CEP is equal to $3\pi/2$ or $\pi$, very similar results as those shown in Fig.~\ref{fig:result}~({\bf c}) and ({\bf d}) are obtained, except that the up and down directions are now reversed. In all these cases, very good agreement is found between experiment and simulation.

\paragraph*{Discussion} We now examine the behavior of the photoelectron distribution in the two cases by using an analytical derivation described in the method section. Assuming two pulses with the same amplitude and a spectral phase difference of $\pi$, the photoionization probability, obtained from Eq.~(\ref{eq:sf1}), is proportional to
\begin{equation}
      |a({\bf p})|^2\propto \sin^2[\pi\Omega/(2\omega)+\eta_p],
\end{equation}
where $\Omega=I_p/\hbar+{\bf p}^2/(2m_e\hbar)$ is the XUV frequency, with $\textbf{p}$ the final momentum of the photoelectron, $m_e$ the electron mass, $\hbar$ the reduced Planck constant and $I_p$ the ionization potential of helium. The phase shift $\eta_p$ is proportional to $\textbf{p}\cdot\textbf{A}_0$ where $\textbf{A}_0$ is the maximum amplitude of the vector potential. The photoelectron distribution in frequency is modulated due to the interference between the electron wave-packets (EWPs) created by the two attosecond pulses. In this case, the interaction with the IR field does not lead to new photoelectron structures (sidebands), but to a shift of the photoelectron peaks. These peaks appear at $\Omega=(2q+1)\omega+2\eta_p\omega/\pi$, which correspond to the position of absorption by odd harmonics, shifted by a quantity proportional to $\eta_p$, thereby depending on the emission direction of the electron as shown in the spectra presented in Fig.~\ref{fig:result}~({\bf c}) and ({\bf e}).

In the case of a main attosecond central pulse and two similar side pulses as in Fig.~\ref{fig:result}~({\bf b}), the photoionization probability is proportional to 
\begin{equation}
   |a({\bf p})|^2\propto 1+4r^2 \cos^2\left(\frac{\pi\Omega}{\omega}\right)-4r \cos\left(\frac{\pi\Omega}{\omega}\right)\cos\left[s(\Omega)+2\eta_p\right],
\end{equation}
where $s(\Omega)$ is the difference in spectral phase and $r$ the amplitude ratio between the side and the central pulses. The second term comes from interference between the first and third EWPs, resulting in peaks at all harmonic frequencies. The third term describes interference between the central EWP with the other two, leading to enhancement or reduction of the sidebands with respect to the main peaks. 
This interpretation, based on the interference of few EWPs, is consistent with a recent theoretical prediction \cite{GramajoPRA2016}. Note that in traditional RABBIT, the spectral phase difference $s(\Omega)$ is very small since it rapidly decreases as the pulse duration increases \cite{GuoJPB2018}, so that the photoelectron distribution remains up-down symmetric. However, in our case, $s(\Omega)$ is not negligible and leads to an up-down asymmetry of the photoelectron emission spectra. 

\begin{figure}[ht]
\includegraphics[width=0.8\textwidth]{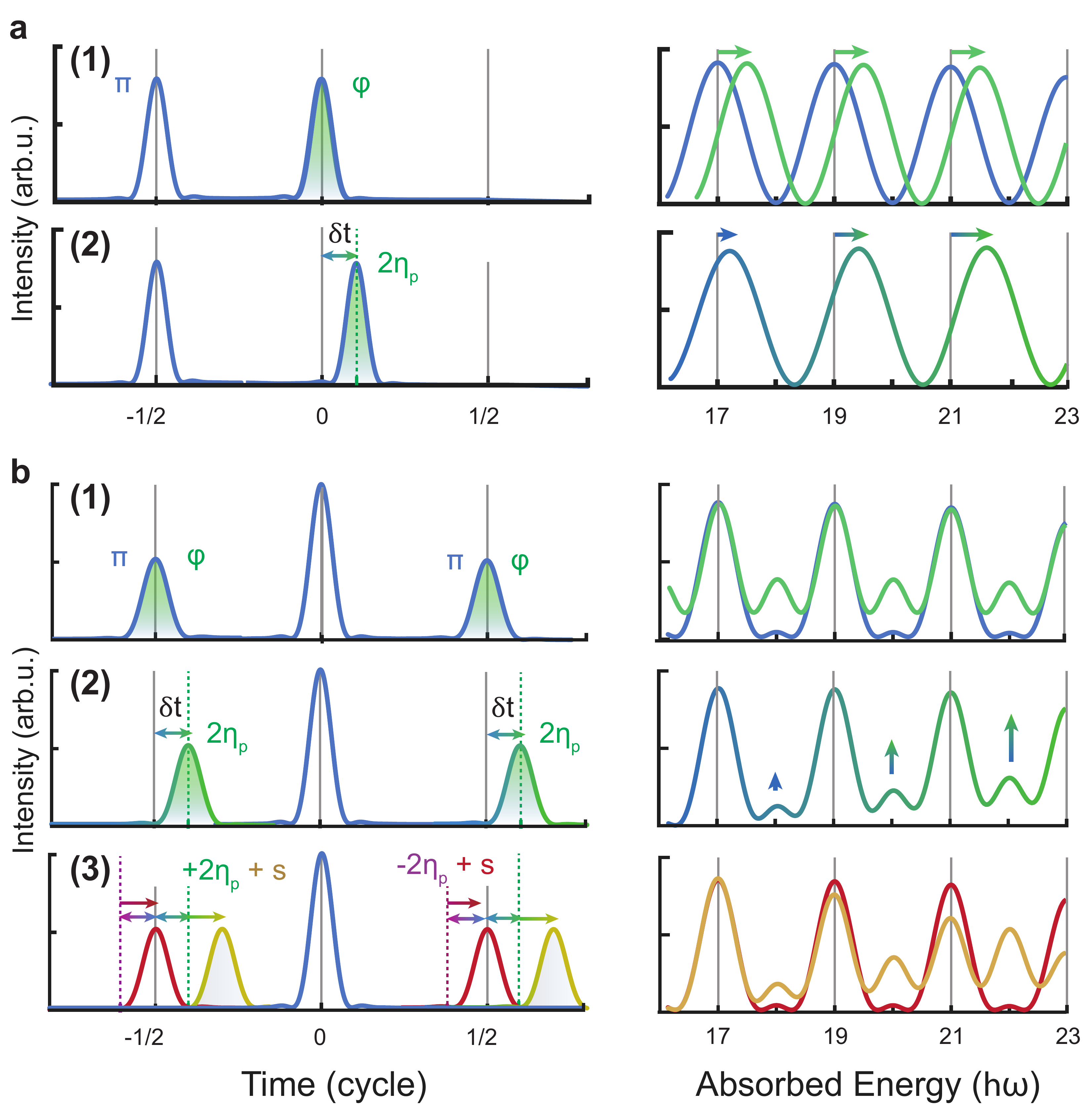}\\
\caption{ \textbf{Interference through multiple temporal slits:} {\bf (a)} The interference of two EWPs separated by half a laser cycle with a $\pi$ phase difference [left plot, panel (1)] leads to a modulation in the energy (frequency) domain, with maxima at the energies corresponding to excitation by odd harmonics [right plot, blue curve, (1)]. A phase change, $\varphi$, of one EWP shifts the interference fringes [green curve, (1)]. A momentum-dependent phase change, $2\eta_p$, (2) leads to an energy-dependent shift of the interference fringes, as well as to a temporal shift ($\delta t$) of one EWP relative to the other. {\bf (b)} The interference of three EWPs separated by half a laser cycle with a $\pi$ phase difference [left plot, panel (1)] leads to interferences with maxima at the energies corresponding to excitation by odd harmonics, and weak ``secondary'' maxima at the SB position [right plot, blue curve, panel (1)]. A phase change between the side and central EWPs ($\varphi$) enhances the SB relative to the main peak [green curve, (1)]. A momentum-dependent phase change ($2\eta_p$) leads to  energy-dependent sideband amplitudes, but no energy shift [right panel (2)]. The spectral phase difference bewteen consecutive attosecond pulses ($s$) enhances (yellow curve) or reduces (red curve) this effect depending on the direction of emission (panel 3).}
\label{fig:slits}
\end{figure}

Finally, we give a simple interpretation of these results based on an analogy with diffraction through two or three slits \cite{LindnerPRL2005,RichterPRL2015}. Fig.~\ref{fig:slits}~({\bf a}) illustrates the two-EWP (or two-slit) case. The Fourier transform of a pair of pulses separated by $\pi/\omega$ leads to a modulation in the frequency domain equal to $2\omega$. When the phase difference between the pulses is $\pi$, constructive interferences take place at  frequencies $\Omega=(2q+1)\omega$, where $q$ is integer, as illustrated in Fig.~\ref{fig:slits}~({\bf a}1). 
An additional constant phase ($\varphi$) imparted in one of the EWPs shifts the interference fringes, as shown by the green curve. In our experiment, the phase difference between the two EWPs, equal to  $2\eta_p$, increases with $|{\bf p}|$, which leads to a (small) time delay ($\delta t\sim$ 100 as) between the two EWPs and a shift increasing with frequency as shown in Fig.~\ref{fig:slits}~({\bf a}2). The sign of the frequency shift depends on the direction of emission of the photoelectron with respect to the polarization, resulting in an asymmetric angular distribution. 

Fig.~\ref{fig:slits}~({\bf b}) illustrates the three-slit case. The Fourier transform of three pulses separated by $\pi/\omega$ and with $\pi$ phase difference leads to interference fringes still separated by $2\omega$ (Fig.~\ref{fig:slits}~({\bf b}1), blue), with a small contribution at frequencies $\Omega=2q\omega$ (sidebands), called ``secondary maxima'' in the theory of diffraction. An additional phase shift ($\varphi$) between consecutive EWPs (green) leads to an enhancement of the sideband peaks. 
In our experiment, the phase difference between consecutive EWPs due to the interaction with the IR field leads to time delays between the EWPs and to sideband intensities increasing with frequency, as shown in Fig.~\ref{fig:slits}~({\bf b}2).
The spectral phase between the side and the central attosecond pulses, $s(\Omega)$, can enhance (compensate for) this effect, increasing (reducing) the sideband intensities, see Fig.~\ref{fig:slits}~({\bf b}3). Since $s(\Omega)+2\eta_p$ depends on the photoelectron emission direction, the angular distribution becomes asymmetric. 
The difference with the two-slit case comes from the fact that the two smaller EWPs have the same phase and amplitude. This fixes the position of the constructive interferences at $\Omega=q\omega$.

The asymmetry in the photoelectron direction of emission is here due to the difference in spectral phase between consecutive attosecond pulses, i.e. the femtosecond chirp of the harmonic emission. This result has a simple interpretation in the spectral domain. The harmonic width becomes broad enough for spectral overlap between the continua reached by two-photon (XUV+IR) and one-photon (XUV) ionization, leading to parity mixing and asymmetric electron emission.

\paragraph*{Conclusion}

The analogy with diffraction by multiple slits allows us to understand the difference between continuous and quantized energy transfer from the IR electromagnetic field to the photoelectron. This difference is due to the temporal confinement of the light-matter interaction, which is reduced to half an optical cycle in this first case while it lasts one optical cycle in the second case. In simple words, the absorption of an energy quantum of light (photon) requires a duration of the light-matter interaction (here limited by the number of attosecond pulses) of at least one optical cycle.  

Our results open a new road in attosecond science, namely, the manipulation of ultrafast processes with a tailored sequence of attosecond pulses, combined with a synchronized weak IR field. From the experimental point-of-view, this achievement is possible thanks to the high CEP-stability and short pulse duration of our laser system. In addition, the high repetition rate of our experiment allows us to measure  three dimensional momentum electron distributions with electron-ion coincidence detection, thus providing a complete kinematic description of the interaction. We envision numerous applications of the time-domain coherent control shown in the present work, e.g. towards two-dimensional spectroscopy of more complex systems at the attosecond temporal resolution and in the XUV spectral range. 

\paragraph*{Methods}

\subparagraph*{Experiment} The experiment was performed with a 200 kHz-repetition rate CEP-stable optical parametric chirped pulse amplification (OPCPA) laser system with 5 $\mathrm{\mu}$J energy per pulse, 820-nm central wavelength and 6 fs pulse duration. The CEP of the laser can be varied with a fused silica wedge pair, as shown in Fig.~2~({\bf a}). The laser pulses are focused using an achromatic lens with 5 cm focal length in an argon gas jet with a 10 bar backing pressure~\cite{HarthJO2017}. 
High-order harmonics are generated, corresponding in the time domain to a train of (primarily) two to three attosecond pulses \cite{GuoJPB2018}. An aluminium filter can be introduced to eliminate the IR field and a concave grating (not shown in Fig.~2) can be inserted after the differential pumping hole in order to disperse the XUV radiation and measure its spectrum with microchannel plate detector. 
The IR field, with an intensity less than 10$^{12}$ W/cm$^2$, and the XUV radiation are focused by a gold-coated toroidal mirror into a vacuum chamber containing an effusive helium gas jet and a 3D momentum spectrometer (see Fig.~\ref{fig:experiment}~({\bf a})). This spectrometer is based on a revised CIEL (``Co\"incidences entre Ions et \'Electrons Localis\'es'') design, 
providing a complete kinematic momentum picture of the emitted ions and electrons without loosing data due to magnetic nodes~\cite{GisselbrechtRSI2005}. The spectrometer orientation is chosen so that the time-of-flight axis coincides with the optical polarization direction. Electron-ion coincidence data are recorded at a typical rate of $\sim$35 kHz, with a negligible amount of false coincidence.

The rotational symmetry of the momentum distribution around the $p_z$-axis (polarization axis) means that the signal can be integrated along the azimuthal angle $\phi$ and subsequently divided by $\sin\theta$, giving the differential cross section. In the XUV-only case, four rings can be identified (see Fig.~\ref{fig:experiment}~(\textbf{b})), corresponding to ionization ($1s \rightarrow \epsilon p$) by absorption of harmonics 17, 19, 21, and 23. 

\subparagraph*{Simulations} The simulations presented in Fig.~\ref{fig:result} ({\bf e}, {\bf f}) have been performed by evaluating the probability amplitude for emission with momentum ${\bf p}$ \cite{QuereJMO2005},  
\begin{equation}
\label{eqn:photo}
a({\bf p})=-i\int_{-\infty}^{\infty} dt~{\bf d}({\bf p})\cdot {\bf E}_\mathrm{XUV}(t)~e^{\frac{i}{\hbar}\left(I_p+\frac{{\bf p}^2}{2m_e}\right)t+i\Phi_\mathrm{IR}({\bf p},t)},
\end{equation}
where $\bf p$ denotes the final electron momentum, $\bf d$ is the dipole moment, ${\bf E}\mathrm{_{XUV}}$ is the XUV field, $m_e$ the electron mass, $\hbar$ the reduced Planck constant and $I_p$ the ionization potential of helium. In the relatively weak field case which is considered in the present work, the action of the laser field reduces to a phase modulation, approximated by 
\begin{equation}
\label{eqn:phase}
    \Phi_\mathrm{IR}({\bf p},t)\approx-\frac{e}{m_e\hbar} \int_t^{+\infty} dt' {\bf p}\cdot\textbf{A}(t'), 
\end{equation}
where $\textbf{A}$ is the vector potential of the IR field. The dipole moment ${\bf d}({\bf p})$ is calculated with an hydrogenic approximation \cite{LewensteinPRA1994}, while both the IR and the XUV fields have been chosen to reproduce the experimental conditions as closely as possible. The XUV attosecond pulses are generated at an IR intensity of $1.1\times 10^{14}$ W/cm$^2$ and  the  IR intensity in the detector chamber is $6\times 10^{11}$ W/cm$^2$. 
For a temporal offset $\tau$ between attosecond pulses and the IR dressing field of $\sim$0.6 optical cycles, excellent agreement between experiment and theory is achieved.    

\subparagraph*{Analytical derivation} The XUV field can be decomposed into a sum of attosecond pulses $E_m(t)$, separated by half the laser period $\pi/\omega$ and centered at $m\pi/\omega$. The XUV and IR fields have the same linear polarization, so we may drop the vector notation unless needed. Assuming that the phase $\Phi_\mathrm{IR}({\bf p},t)$ does not vary much over the duration of the attosecond pulse, using $\textbf{A}(t)=\textbf{A}_0\cos[\omega (t-\tau)]$, Eq.~(\ref{eqn:phase}) becomes:  
\begin{equation}
    \Phi_\mathrm{IR} \left({\bf p},\frac{m\pi}{\omega}\right)=-\frac{e{\bf p}\cdot{\bf A}_0 }{m_e\hbar\omega} (-1)^m \sin(\omega\tau),
\end{equation} 
which we write in a more compact form as $-(-1)^m \eta_p$. Introducing the spectral amplitude $E_m(\Omega)$, equal to the Fourier transform of $E_m(t)$, where $\Omega=I_p/\hbar+{\bf p}^2/(2m_e\hbar)$ is the XUV frequency, Eq.~(\ref{eqn:photo}) simplifies to: 
\begin{equation}
a({\bf p}) \approx -i d({\bf p})\sum_m e^{-i(-1)^m \eta_p} e^{\frac{im\pi\Omega }{\omega}} E_m(\Omega).
\label{eq:sf1}
\end{equation}
In the perturbative limit ($\eta_p\ll1$), Eq.~(\ref{eq:sf1}) can be written as the sum of two terms. The first term describes ionization by absorption of one photon,
\begin{equation}
a_1({\bf p}) \approx -id({\bf p}) \sum_m  e^{i\frac{m\pi\Omega}{\omega}} E_m(\Omega).
\label{eq:a1}
\end{equation}
When consecutive attosecond pulses have a phase difference of $\pi$, $a_1({\bf p})$ is maximum when $\Omega= (2q+1) \omega$, where $q$ is integer, corresponding to ionization by absorption of odd-order harmonics of the laser field.  
The second term includes the interaction with the IR field, 
\begin{equation}
a_2({\bf p}) = -\eta_p   d({\bf p}) \sum_m (-1)^m e^{\frac{im\pi\Omega}{\omega}} E_m(\Omega).
\label{eq:a2}
\end{equation}
When attosecond pulses have approximately the same amplitudes and a phase difference between consecutive pulses of $\pi$, $a_2({\bf p})$ is maximum when $\Omega= 2q \omega$, where $q$ is integer, leading thus to sideband peaks in the photoelectron distribution, at energies that would correspond to ionization by absorption of even-order harmonics. To get the photoelectron distribution the two amplitudes $a_1({\bf p})$ and $a_2({\bf p})$ are added coherently. An up-down asymmetric photoelectron spectrum requires that the two terms overlap spectrally, allowing mixing of states with different parity.

\bibliography{Ref_lib.bib}

\begin{thebibliography}{10}

\bibitem{PlanckAdP1901}
Planck, M.
\newblock Ueber das gesetz der energieverteilung im normalspectrum.
\newblock {\em Ann. Phys.} {\bf 309}, 1  (1901).

\bibitem{EinsteinAdP1905}
Einstein, A.
\newblock {\"U}ber einen die erzeugung und verwandlung des lichtes betreffenden
  heuristischen gesichtspunkt.
\newblock {\em Ann. Phys.} {\bf 322}, 132--148  (1905).

\bibitem{LaporteJOSA1925}
Laporte, O., Meggers, W.~F.
\newblock Some rules of spectral structure.
\newblock {\em J. Opt. Soc. Am. A} {\bf 11}, 459--463  (1925).

\bibitem{YangPR1948}
Yang, C.~N.
\newblock On the angular distribution in nuclear reactions and coincidence
  measurements.
\newblock {\em Phys. Rev.} {\bf 74}, 764  (1948).

\bibitem{CooperJCP1968}
Cooper, J., Zare, R.~N.
\newblock Angular {D}istribution of {P}hotoelectrons.
\newblock {\em J. {C}hem. {P}hys} {\bf 48}, 942  (1968).

\bibitem{MaimanNature1960}
Maiman, T.~H.
\newblock Stimulated optical radiation in ruby.
\newblock {\em Nature} {\bf \textbf{187}}, 493  (1960).

\bibitem{RowePA1973}
Rowe, E.~M., Mills, F.~E.
\newblock {Tantalus. 1. A Dedicated Storage Ring Synchrotron Radiation source}.
\newblock {\em Part. Accel.} {\bf 4}, 211-227  (1973).

\bibitem{QuackHandbook2011}
Quack, M., Merkt, F.
\newblock {\em Handbook of high-resolution spectroscopy}.
\newblock Wiley-Blackwell (2011).

\bibitem{BeckerVUV2012}
Becker, U., Shirley, D.~A.
\newblock {\em VUV and Soft X-ray Photoionization}.
\newblock Springer Science \& Business Media (2012).

\bibitem{AgostiniPRL1979}
Agostini, P., Fabre, F., Mainfray, G., Petite, G., Rahman, N.~K.
\newblock Free-{F}ree {T}ransitions {F}ollowing {S}ix-{P}hoton {I}onization of
  {X}enon {A}toms.
\newblock {\em Phys. {R}ev. {L}ett.} {\bf 42}, 1127  (1979).

\bibitem{BucksbaumPRL1987}
Bucksbaum, P.~H., Bashkansky, M., McIlrath, T.~J.
\newblock Scattering of {E}lectrons by {I}ntense {C}oherent {L}ight.
\newblock {\em Phys. {R}ev. {L}ett.} {\bf 58}, 349  (1987).

\bibitem{YinPRL1992}
Yin, Y., Chen, C., Elliott, D.~S., Smith, A.~V.
\newblock Asymmetric {P}hotoelectron {A}ngular {D}istributions from
  {I}nterfering {P}hotoionization {P}rocesses.
\newblock {\em Phys. {R}ev. {L}ett.} {\bf 69}, 2353  (1992).

\bibitem{LaurentPRL2012}
Laurent, G., et~al.
\newblock Attosecond control of orbital parity mix interferences and the
  relative phase of even and odd harmonics in an attosecond pulse train.
\newblock {\em Phys. Rev. Lett.} {\bf 109}, 083001  (2012).

\bibitem{PaulusNature2001}
Paulus, G.~G., et~al.
\newblock Absolute-phase phenomena in photoionization with few-cycle laser
  pulses.
\newblock {\em Nature} {\bf 414}, 182  (2001).

\bibitem{PaulusPRL2003}
Paulus, G.~G., et~al.
\newblock Measurement of the phase of few-cycle laser pulses.
\newblock {\em Phys. {R}ev. {L}ett.} {\bf 91}, 253004  (2003).

\bibitem{McPhersonJOSAB1987}
McPherson, A., et~al.
\newblock Studies of multiphoton production of vacuum-ultraviolet radiation in
  the rare gases.
\newblock {\em J. {O}pt. {S}oc. {A}m.~{B}} {\bf \textbf{4}}, 595  (1987).

\bibitem{FerrayJPB1988}
Ferray, M., et~al.
\newblock Multiple-harmonic conversion of 1064 nm radiation in rare gases.
\newblock {\em J. {P}hys. {B}} {\bf \textbf{21}}, L31  (1988).

\bibitem{SchultzeScience2010}
Schultze, M., et~al.
\newblock Delay in photoemission.
\newblock {\em Science} {\bf 328}, 1658-1662  (2010).

\bibitem{KlunderPRL2011}
Kl\"under, K., et~al.
\newblock Probing single-photon ionization on the attosecond time scale.
\newblock {\em Phys. Rev. Lett.} {\bf 106}, 143002  (2011).

\bibitem{OssianderNatPhys2017}
Ossiander, M., et~al.
\newblock Attosecond correlation dynamics.
\newblock {\em Nature Physics} {\bf 13}, 280 - 285  (2017).

\bibitem{IsingerScience2017}
Isinger, M., et~al.
\newblock Photoionization in the time and frequency domain.
\newblock {\em Science} {\bf 358}, 893--896  (2017).

\bibitem{HuppertPRL2016}
Huppert, M., Jordan, I., Baykusheva, D., Conta, A., W\"orner, H.~J.
\newblock Attosecond delays in molecular photoionization.
\newblock {\em Phys. Rev. Lett.} {\bf 117}, 093001  (2016).

\bibitem{CattaneoNatPhys2018}
Cattaneo, L., et~al.
\newblock Attosecond coupled electron and nuclear dynamics in dissociative
  ionization of h-2.  (2018).

\bibitem{VosScience2018}
Vos, J., et~al.
\newblock Orientation-dependent stereo wigner time delay and electron
  localization in a small molecule.
\newblock {\em Science} {\bf 360}, 1326 - 1330  (2018).

\bibitem{CavalieriNature2007}
Cavalieri, A.~L., et~al.
\newblock Attosecond spectroscopy in condensed matter.
\newblock {\em Nature} {\bf 449}, 1029  (2007).

\bibitem{LucchiniScience2016}
Lucchini, M., et~al.
\newblock Attosecond dynamical franz-keldysh effect in polycrystalline diamond.
\newblock {\em Science} {\bf 353}, 916--919  (2016).

\bibitem{KienbergerScience2002}
Kienberger, R., et~al.
\newblock Steering {A}ttosecond {E}lectron {W}ave {P}ackets with {L}ight.
\newblock {\em Science} {\bf \textbf{297}}, 1144  (2002).

\bibitem{PazourekRMP2015}
Pazourek, R., Nagele, S., Burgd\"orfer, J.
\newblock Attosecond chronoscopy of photoemission.
\newblock {\em Rev. Mod. Phys.} {\bf 87}, 765--802  (2015).

\bibitem{VeniardPRA1996}
V\'eniard, V., Ta\"ieb, R., Maquet, A.
\newblock Phase dependence of ({N}+1)\,-\,color ({N}$>$1) ir-uv photoionization
  of atoms with higher harmonics.
\newblock {\em Phys. {R}ev.~{A}} {\bf \textbf{54}}, 721  (1996).

\bibitem{PaulScience2001}
Paul, P., et~al.
\newblock Observation of a train of attosecond pulses from high harmonic
  generation.
\newblock {\em Science} {\bf 292}, 1689  (2001).

\bibitem{MullerAPB2002}
Muller, H.
\newblock Reconstruction of attosecond harmonic beating by interference of
  two-photon transitions.
\newblock {\em Appl. {P}hys.~{B}} {\bf \textbf{74}}, 17  (2002).

\bibitem{HoglePRL2015}
Hogle, C., et~al.
\newblock Attosecond coherent control of single and double photoionization in
  argon.
\newblock {\em Phys. {R}ev. {L}ett.} {\bf 115}, 173004  (2015).

\bibitem{HeuserPRA2016}
Heuser, S., et~al.
\newblock Angular dependence of photoemission time delay in helium.
\newblock {\em Phys. {R}ev. {A}} {\bf 94}, 063409  (2016).

\bibitem{GuoJPB2018}
Guo, C., et~al.
\newblock Phase control of attosecond pulses in a train.
\newblock {\em J. Phys. B} {\bf 51}, 034006  (2018).

\bibitem{DornerPhysRep2000}
Dorner, R., et~al.
\newblock Cold {T}arget {R}ecoil {I}on {M}omentum {S}pectroscopy: a 'momentum
  microscope' to view atomic collision dynamics.
\newblock {\em Physics {R}eports} {\bf 330}, 95-192  (2000).

\bibitem{Ullrich03rpp}
Ullrich, J., et~al.
\newblock Recoil-ion and electron momentum spectroscopy: reaction-microscopes.
\newblock {\em Reports on Progress in Physics} {\bf 66}, 1463-1545  (2003).

\bibitem{LewensteinPRA1994}
Lewenstein, M., Balcou, P., Ivanov, M., L'Huillier, A., Corkum, P.
\newblock Theory of high-order harmonic generation by low-frequency laser
  fields.
\newblock {\em Phys. {R}ev. {A}} {\bf 49}, 2117  (1994).

\bibitem{QuereJMO2005}
Qu\'er\'e, F., Mairesse, Y., Itatani, J.
\newblock Temporal characterization of attosecond {XUV} fields.
\newblock {\em J. {M}od. {O}pt.} {\bf 52}, 339  (2005).

\bibitem{GramajoPRA2016}
Gramajo, A.~A., R.~Della~Picca, R., Garibotti, C.~R., Arb\'o, D.~G.
\newblock Intra- and intercycle interference of electron emissions in
  laser-assisted xuv atomic ionization.
\newblock {\em Phys. {R}ev. {A}} {\bf 94}, 053404  (2016).

\bibitem{LindnerPRL2005}
Lindner, F., et~al.
\newblock Attosecond {D}ouble-{S}lit {E}xperiment.
\newblock {\em Phys. {R}ev. {L}ett.} {\bf 95}, 040401  (2005).

\bibitem{RichterPRL2015}
Richter, M., et~al.
\newblock Streaking temporal double-slit interference by an orthogonal
  two-color laser field.
\newblock {\em Phys. Rev. Lett.} {\bf 114}, 143001  (2015).

\bibitem{HarthJO2017}
Harth, A., et~al.
\newblock Compact 200 {kHz} {HHG} source driven by a few-cycle {OPCPA}.
\newblock {\em J. Opt.} {\bf 20}, 014007  (2017).

\bibitem{GisselbrechtRSI2005}
Gisselbrecht, M., Huetz, A., Lavollée, M., Reddish, T.~J., Seccombe, D.~P.
\newblock Optimization of momentum imaging systems using electric and magnetic
  fields.
\newblock {\em Rev. Sci. Instrum.} {\bf 76}, 013105  (2005).

\end{thebibliography}
\bibliographystyle{myunsrt_NC}

\section*{Acknowledgments}

The authors thank Marcus Dahlstr\"om, David Busto, Ivan Sytcevich and Fabian Langer for insightful scientific discussions. The authors acknowledge support from the Swedish Research Council, the European Research Council (advanced grant PALP-339253) and the Knut and Alice Wallenberg Foundation.  

\section*{Authors contribution}
Y-C.C., S.M., S.N. and M.G. have carried out the experiments. Y-C.C. and L.R. performed the analysis of the data. S.M. and S.C. have carried out simulations. These two authors, Y-C.C. and S.M., have equally contributed
to the article. Y-C.C., S.N. and M.G. have provided the main experimental setup. S.M., C.G., A.H., M.M. and C.A. have provided an advanced 200kHz laser system. M.G. and A.L. have contributed to the theoretical ideas. A.L. performed the analytical derivation. Y-C.C., S.M., J.V. and C.A. have contributed to the writing of the
manuscript. M.G. and A.L. have written the main part of the manuscript.
\clearpage


\newpage

\end{document}